# A calculation scheme for spectral densities of strongly correlated electron systems using exact diagonalization of dynamical mean field theory

Song-Jin O, Hak-Chol Pak, Kwang-Il Ryom, and Song-Jin Im

*Department of Physics, Kim Il Sung University, Pyongyang, Democratic People's Republic of Korea*

**Abstract**

A new approach for calculating spectral density functions of strongly correlated electron systems is proposed within the exact diagonalization method of dynamical mean-field theory (DMFT). This approach is based on the analytic continuation of spectral density function obtained by conventional exact diagonalization method of DMFT and its results are more reasonable in shape.

As an example of its application, the Mott transition in a strongly correlated electron system is investigated using this new approach.

## 1. Introduction

The dynamic mean-field theory (DMFT) has been known today as a powerful method to interpret various physical properties of strongly correlated electron systems.[1] Quantum Monte Carlo method (QMC) and exact diagnolization method (ED) are recognized as the most exact method in solving the DMFT equations.[2] While QMC[3] involves a vast amount of calculation due to long interval of integral in low temperature regime, the ED method is more accurate in this regime. Furthermore it has been reported recently that the calculation with a small number of bath sites can also achieve high accuracy when there are many correlation orbits or sites in a unit cell, and so the ED method is widely being used.[4] In this background, we are going to deal with strongly correlated electron system by using ED.

The ED method of DMFT has one shortcoming associated with spectral density function. In the ED method, lattice system is approximated as central lattice site with several bath sites around it, so one-particle spectral density function obtained by ED method consists of several sharp peaks.[5] Especially in weak correlation limit, spectral density function must be the same as the density of states determined by energy band structure. However, spectral density function obtained by ED method still has sharp peaks. This motivated us to propose a new approach to calculate one-particle spectral density function.

## 2. New algorithm for calculating spectral density based on exact diagonalization method of dynamical mean field theory

As mentioned before, in ED method, strongly correlated lattice systems are approximated as interacting central impurity lattice site and several bath sites around it.

Retarded Green function for impurity sites is defined as

$$\tilde{G}^R_{\alpha\beta}(t) \equiv -i\theta(t)\text{Tr}\{\hat{\rho}_G[\hat{d}^H_\alpha(t)\hat{d}^\dagger_\beta + \hat{d}^\dagger_\beta \hat{d}^H_\alpha(t)]\} \qquad (1)$$

where $\hat{d}_\alpha$, $\hat{d}^\dagger_\beta$ destroys and creates a d-orbital electron with spin $\alpha$, $\beta$ on the impurity site respectively, and $\hat{d}^H_\alpha(t) \equiv e^{i\hat{K}t}\hat{d}_\alpha e^{-i\hat{K}t}$ is Heisenberg representation with grand canonical Hamiltonian $\hat{K} = \hat{H} - \mu\hat{N}$.

When there is no external magnetic field, the retarded impurity Green function has the form of $\tilde{G}^R_{\alpha\beta}(t) = \tilde{G}^R(t)\delta_{\alpha\beta}$ in the paramagnetic phase. After evaluating the retarded impurity Green function, we perform Fourier transformation of it to obtain

$$\tilde{G}^R(\omega) = \frac{1}{Z} \sum_{i,j} (e^{-\beta E_i} + e^{-\beta E_j}) |\langle i|\hat{d}^\dagger_\uparrow|j\rangle|^2 \cdot \frac{1}{\omega + i\delta + E_j - E_i}. \quad (2)$$

where $\delta = 0^+$.

We can also obtain Fourier transformation of temperature Green function for impurity sites in a similar way:

$$\tilde{G}(i\omega_n) = \frac{1}{Z} \sum_{i,j} (e^{-\beta E_i} + e^{-\beta E_j}) |\langle i|\hat{d}^\dagger_\uparrow|j\rangle|^2 \cdot \frac{1}{i\omega_n + E_j - E_i}$$

Comparing Eq. (2) with above equation, it can be seen that if we replace $i\omega_n$ in $\tilde{G}(i\omega_n)$ with $\omega + i\delta$ then we obtain $\tilde{G}^R(\omega)$. In other words, one can obtain $\tilde{G}^R(\omega)$ by analytic continuation of $\tilde{G}(i\omega_n)$ into upper half of complex plane.

On the other hand, one-particle spectral density function can be written as follows.

$$A(\omega) \equiv -\frac{1}{\pi} \operatorname{Im} \tilde{G}^R(\omega) = \frac{1}{Z} \sum_{i,j} (e^{-\beta E_i} + e^{-\beta E_j}) |\langle i|\hat{d}^\dagger_\uparrow|j\rangle|^2 \delta(\omega + E_j - E_i) \quad (3)$$

According to the general theory of Green function, one-particle spectral density function gives density of states determined by the energy band for non-interacting lattice system, i.e.

$$A(\omega) = D(\omega + \mu) \quad \text{when} \quad U = 0.$$

In early method to calculate one-particle spectral density function, after self-consistent calculation of DMFT, one calculates impurity retarded Green function $\tilde{G}^R(\omega)$ according to Eq. (2) using $E_i$, $|i\rangle$ obtained by ED. Then as in Eq. (3), spectral density function $A(\omega)$ was obtained from the imaginary part of $\tilde{G}^R(\omega)$. In such processes the number of bath sites cannot be so large due to limited computation power and usually takes a value between 4 and 9. In this case spectral density function $A(\omega)$ determined by above method consists of several sharp peaks. Even for non-interacting system, $A(\omega)$ still has sharp peaks.

In order to avoid this problem we propose a new algorithm to compute one-particle spectral density function $A(\omega)$. This algorithm can be summarized as follows.

Firstly, evaluate the retarded impurity Green function $\tilde{G}^R(\omega)$ by Eq. (2) after self-consistent calculation.

Secondly, evaluate retarded propagator.

$$g^R(\omega) = \frac{1}{\omega + i\delta + \mu - \sum_{p=2}^{n_s} \frac{V_p^2}{\omega + i\delta - \tilde{\varepsilon}_p}} \quad (4)$$

This is a variant of mean-field propagator

$$g^{n_s}(i\omega_n) = \frac{1}{i\omega_n + \mu - \sum_{p=2}^{n_s} \frac{V_p^2}{i\omega_n - \tilde{\varepsilon}_p}} \quad (5)$$

, where $i\omega_n$ is replaced by $\omega + i\delta$. In other words, it is the analytic continuation of $g^{n_s}(i\omega_n)$ into upper half of complex plane.

Thirdly, evaluate retarded self-energy.

$$\Sigma^R(\omega) = [g^R(\omega)]^{-1} - [\tilde{G}^R(\omega)]^{-1} \quad (6)$$

Retarded self-energy $\Sigma^R(\omega)$ is the analytic continuation of self-energy of temperature Green function

$$\Sigma(i\omega_n) = [g^{n_s}(i\omega_n)]^{-1} - [\tilde{G}(i\omega_n)]^{-1} \quad (7)$$

into upper half of complex plane.

Then evaluate retarded lattice Green function by the following equation.

$$G^R(\omega) = \frac{1}{N}\sum_{\mathbf{k}}\frac{1}{\omega + i\delta - \varepsilon_{\mathbf{k}} + \mu - \Sigma^R(\omega)} = \int_{-\infty}^{\infty}d\varepsilon \frac{D(\varepsilon)}{\omega + i\delta + \mu - \Sigma^R(\omega) - \varepsilon} \qquad (8)$$

This correponds to the analytic continuation of lattice Green function $G(i\omega_n)$ into the upper half of complex plane.

Finally, evaluate the one-particle spectral density function by the following equation.

$$A(\omega) = -(1/\pi)\operatorname{Im}G^R(\omega) \qquad (9)$$

Spectral density obtained this way approaches exactly to the density of states for the energy band in the weak interaction limit.

### 3. Mott transition in strongly correlated electron systems

ED of DMFT has advantage of simple and straightforward calculation, so we can construct a complete program system once we build an algorithm to calculate spectral density based on the method. We evaluate one-particle spectral density function of single-band Hubbard model by using our new algorithm to calculate spectral densities and investigate Mott transition of strongly correlated electron systems.

In this paper, we limit our consideration only to the case of two dimensional square lattice with only nearest hopping. Then the Hamiltonian of this square lattice is

$$\hat{H} = -t\sum_{<ij>,\sigma}(\hat{c}_{i\sigma}^+\hat{c}_{j\sigma} + \hat{c}_{j\sigma}^+\hat{c}_{i\sigma}) + \sum_{i,\sigma}(\varepsilon_d - \mu)\hat{c}_{i\sigma}^+\hat{c}_{i\sigma} + U\sum_i \hat{n}_{i\uparrow}\hat{n}_{i\downarrow} \qquad (10)$$

, where the sum $\sum_{<ij>}$ is carried out for all pairs of nearest sites, $t$ is the nearest hopping parameter, $U$ is the on-site Coulomb repulsion, $\varepsilon_d$ is the energy level of d-orbital and $\mu$ is the chemical potential.

If we set zero of energy as the d-orbital level $\varepsilon_d$, then Hamiltonian (10) can be written as follows because $\mu = U/2$ for the half-filled Hubbard model.

$$\hat{H} = -t\sum_{<ij>,\sigma}(\hat{c}_{i\sigma}^+\hat{c}_{j\sigma} + \hat{c}_{j\sigma}^+\hat{c}_{i\sigma}) + U\sum_i(\hat{n}_{i\uparrow} - \frac{1}{2})(\hat{n}_{i\downarrow} - \frac{1}{2}) - \frac{N}{4}U \qquad (11)$$

We introduce Fourier transformation of the kinetic energy in Eq. (11) to obtain

$$\hat{H} = \sum_{\mathbf{k},\sigma}\varepsilon(\mathbf{k})\hat{c}_{\mathbf{k}\sigma}^+\hat{c}_{\mathbf{k}\sigma} + U\sum_i(\hat{n}_{i\uparrow} - \frac{1}{2})(\hat{n}_{i\downarrow} - \frac{1}{2}) - \frac{N}{4}U \qquad (12)$$

$$\varepsilon(\mathbf{k}) = -2t(\cos k_x + \cos k_y) \qquad (13)$$

, where $\mathbf{k}$ is the two dimensional wave vector and the lattice constant was set as unity.

For numerical evaluation we set the hopping parameter as $t = 0.25\text{eV}$ in which case energy band width is $W = 8t = 2.00\text{eV}$. In the ED, the number of sites in the finite lattice system $n_s$ was chosen as 6. (number of bath sites is $n_b = n_s - 1 = 5$)

Density of states corresponding to the dispersion relation of Eq. (13) is shown as solid line in Fig. 1. Spectral densities calculated by early and new algorithm are shown as dotted and dashed lines respectively in non-interacting case.

As can be seen from Fig. 1, our spectral density nearly coincides with density of states. Slight differences are due to using finite positive imaginary part of frequency ($\delta = 0.05t$) rather than $\delta = 0^+$ when evaluating retarded Green function. On the contrary spectral density calculated by early method greatly differs from density of states in spite of no interaction and consists of several sharp peaks of delta function.

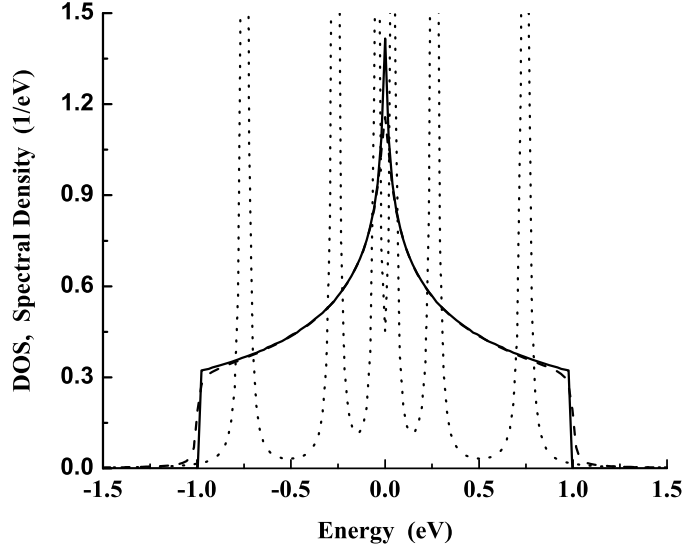

FIG. 1. Density of states and spectral densities of 2D square lattice with only nearest hopping.
The solid line is the density of states determined from the energy band.
The dotted line is the spectral density from early algorithm and
the dashed line is one from our new algorithm.

We calculated spectral density function with gradually increasing Coulomb repulsion $U$ from zero at the fixed temperature $kT = t/16 = 15.625 \text{meV}$.

Fig. 2 shows the change of spectral density when the Coulomb repulsion $U$ is gradually increased. It can be seen from the figure that the system undergoes phase transition from metal to insulator when Coulomb repulsion increases from $U = 9.00t$ to $U = 10.00t$. But the result from early algorithm (dotted line) has sharp peaks for overall region and one can hardly determine the point of phase transition from it.

For weak repulsion the spectral density determined by our new algorithm has the shape of a main peak with the small peaks overlapped on it. However, when the system changes to strongly correlated system, small peaks disappear to form several clearly distinguishable main peaks. Especially when transition from metal to insulator takes place, quasi-particle peak in the middle of five-peaks structure splits into two small peaks to form six-peaks structure. Then these two peaks vanish and four-peaks structure is formed.

Fig. 3 shows the change of spectral density in the vicinity of Mott transition. It can be seen from the figure that the Mott transition occurs at $U_c = 9.58t$. And one can see that the Mott transition is a first order transition from the fact that the spectral density has a sudden change when the parameter $U$ changes from $U = 9.57t$ to $U = 9.58t$. This is in good agreement with previous result [6].

Through this calculation we conclude that our new algorithm for calculating spectral density provides us with exact result in the limit of weak correlation and gives us a reasonable description of the experiment of Mott transition.

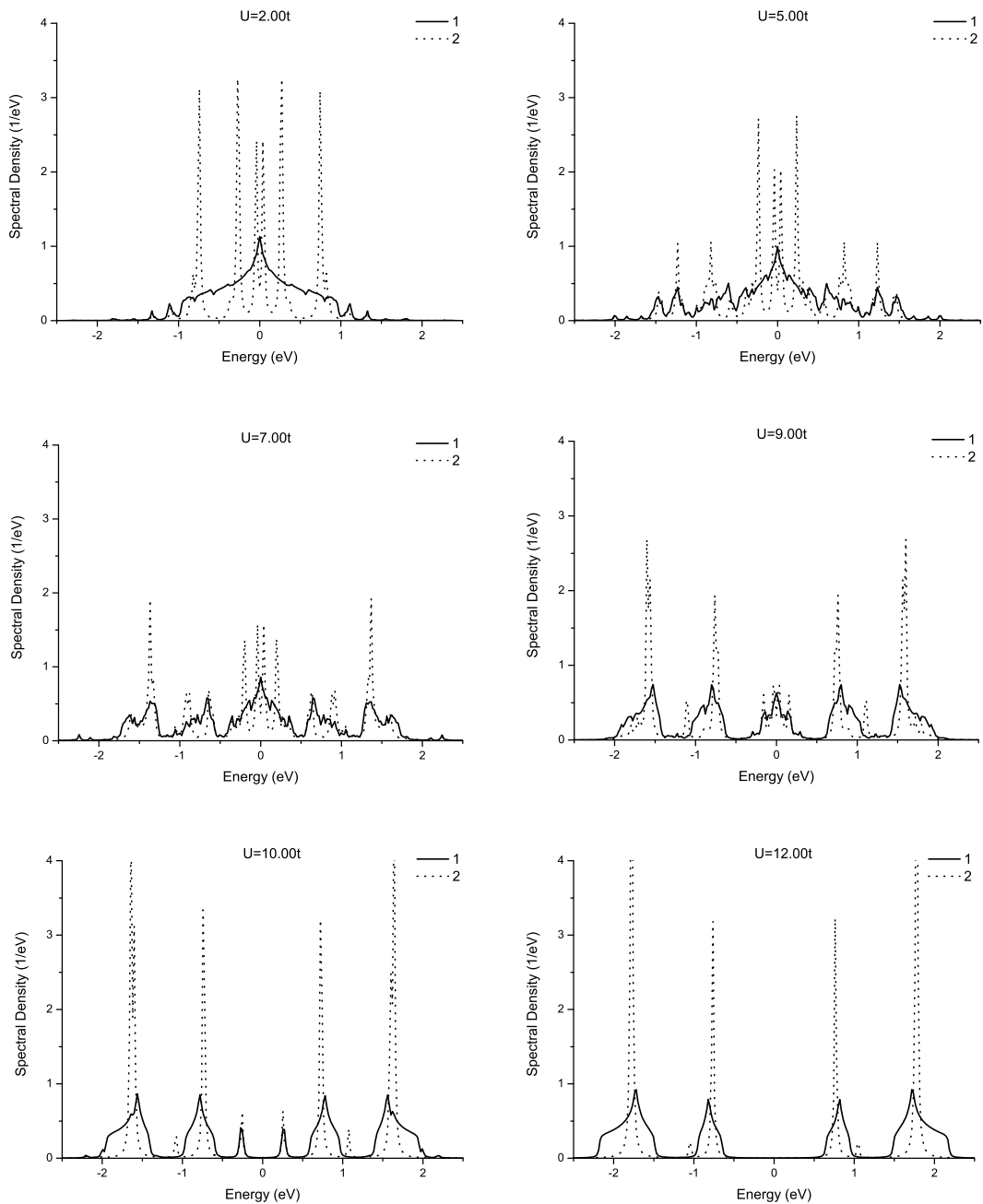

FIG. 2. Change of spectral density function when Coulomb repulsion is increased ( $k_BT = t/16$ ). The solid line is the result from our new algorithm ( $\rho(E) = -(1/\pi)\operatorname{Im} G^R(E)$ ) and the dotted line is one from early algorithm ( $\tilde{\rho}(E) = -(1/\pi)\operatorname{Im} \tilde{G}^R(E)$ ).

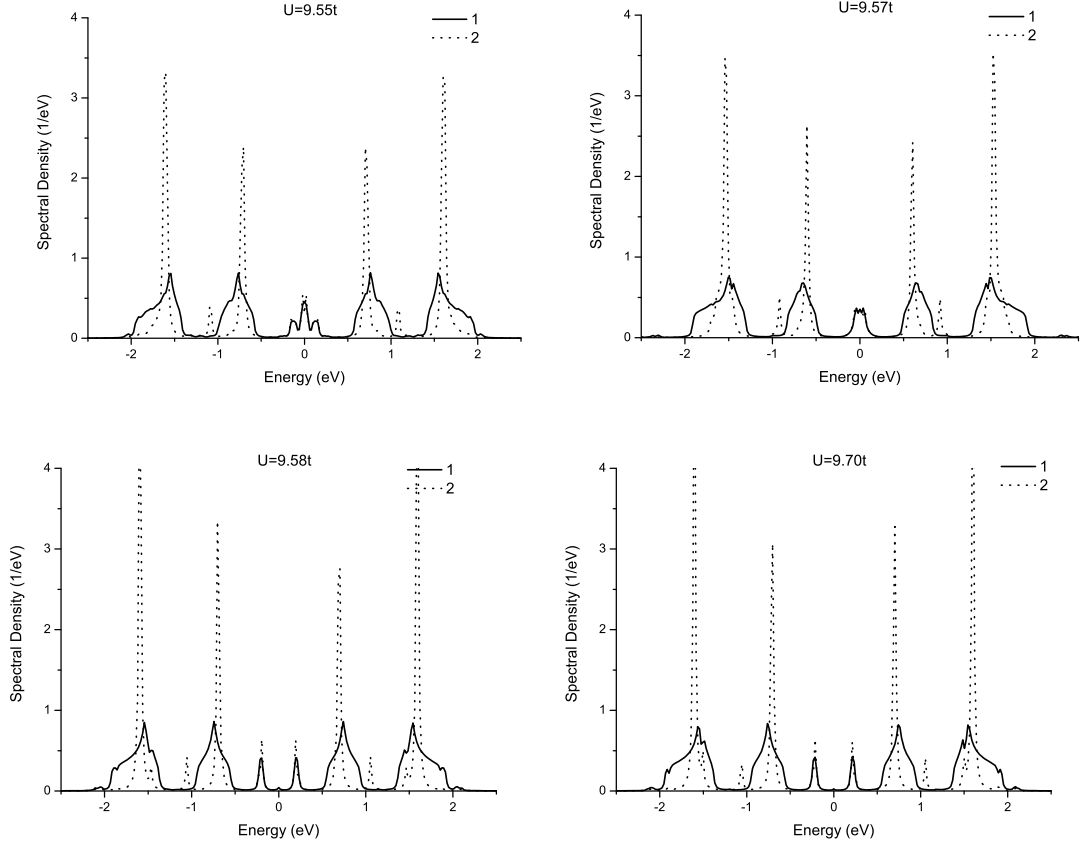

FIG. 3. Change of spectral density function in the vicinity of Mott transition ( $k_B T = t/16$ ). The solid and dotted lines have the same meaning as in Fig. 2.

## 4. Summary

A new approach for calculating spectral density functions of interacting electron systems is proposed based on exact diagonalization method of DMFT. This new approach provides us with exact result in the limit of weak correlation and more reliable results than earlier even in the limit of strong correlation.

Through the calculation of the spectral density function we have intuitively shown the Mott transition from conductor to insulator with increasing Coulomb repulsion. With our new approach to calculate spectral density one can find the point of the Mott transition more precisely than earlier method.